\documentclass[a4paper,11pt]{article}
\usepackage[dvips]{graphicx}
\usepackage{amssymb}
\usepackage{amsmath}

\usepackage{cite}

\begin{document}

\title{F-theory and the Witten Index}
\author{V.K.Oikonomou\thanks{
voiko@physics.auth.gr}\\
Technological Education Institute of Serres, \\
Department of Informatics and Communications 62124 Serres, Greece\\
and\\
Department of Theoretical Physics Aristotle University of Thessaloniki,\\
Thessaloniki 541 24 Greece} \maketitle

\begin{abstract}
We connect the fermionic fields, localized on the intersection
curve $\Sigma$ of two D7-branes with zero background flux, to a
$N=2$ supersymmetric quantum mechanics algebra, within the
theoretical framework of F-theory.
\end{abstract}

\section*{Introduction}

F-theory
\cite{vafa1,origins1,origins2,vafa2,vafa3,review1,review2,review3,review4,uplift,f1,f2,f3,f4,f5,f6,f7,f8,f9,f10,f11,f12,f13,f14,f15,f16,f17,f18,f19,f21,f22,f23,f24,f25,f26,f27,f28,f29,f30,f31,f32,f33,f34,f35,f36,f37,f38,f39,f40,f41,f42,f43,f44,f45,f46,f47,f48,f49,f50,f51,f52,f53,f54,f55,f56,f57,f57a,f58b,f58,f59,f60,f61,f62,f63,f64,f65,f66,f67,f68},
is a strongly coupled formulation of type IIB superstring theory
with 7-branes. The most interesting achievement is that it
provides a consistent non-perturbative theoretical apparatus for
consistent GUT model building
\cite{vafa2,vafa3,review1,review2,review3,review4,uplift,f1,f2,f3,f4,f5,f6,f7,f8,f9,f10,f11,f12,f13,f14,f15,f16,f17,f18,f19,f21,f22,f23,f24,f25,f26,f27,f28,f29,f30,f31,f32,f33,f34,f35,f36,f37,f38,f39,f40,f41,f42,f43,f44,f45,f46,f47,f48,f49,f50,f51,f52,f53,f54,f55,f56,f57,f57a,f58b,f58,f59,f60,f61},
with gravity being decoupled from the theory. The fields of the
Standard Model arise from local F-theory GUT geometries which
constrain the phenomenological outcomes of the models.

Perturbative local D-branes models offer the possibility of
generating the Standard Model fields existing on branes, in the
absence of gravity (see e.g., references in \cite{review4}). For
example in type IIB, GUT groups like $SO(10)$ or $SU(5)$ can be
obtained on a stack of D7-branes or their orientifolds. However
couplings such as $5_H\times 10_M\times 10_M$ for $SU(5)$, or
representations like the spinor $16$ of $SO(10)$, cannot be
realized within the perturbative framework of superstring
theories. F-theory provides a consistent theoretical framework
(actually a UV completion of type IIB) on which GUT theories can
be built.

Realistic GUT models must necessarily include mechanisms that
break the GUT group, to the Standard Model one. In addition,
proton stability must be ensured, and the various flavor mass
hierarchies must be explained. Furthermore a coupling to a
supersymmetry breaking sector must be included. F-theory
compactifications deal with all the aforementioned problems and
offers phenomenologically interesting, geometry based answers.
Moreover the geometry of the compactifications provides rich
alternatives in model building and different approaches to the
same problem\cite{vafa1,vafa2,review1,review2,review3,review4}.

The most interesting F-theory compactifications are those which
preserve $N=1$ supersymmetry in four dimensions
\cite{vafa2,vafa3}. This condition is met for compactifications on
a complex dimension four Calabi-Yau manifold. Nevertheless the six
physical internal dimensions are a threefold, say $B_3$,  that
needs not to be Calabi-Yau.

D7-branes naturally arise when Calabi-Yau fourfolds are
considered. Actually the locations of the seven branes are given
by the roots of the discriminant of a complex dimensional equation
in $B_3$ \cite{vafa2,review2,review3,review4,uplift}. The
singularities in turn, are classified according to the Kodaira
classification of singular fibers (ADE singularities). It is this
classification that realizes gauge groups.

Since the originating papers on F-theory
\cite{vafa1,origins1,origins2}, a vast amount of work has been
carried out
\cite{vafa2,vafa3,review1,review2,review3,review4,uplift,f1,f2,f3,f4,f5,f6,f7,f8,f9,f10,f11,f12,f13,f14,f15,f16,f17,f18,f19,f21,f22,f23,f24,f25,f26,f27,f28,f29,f30,f31,f32,f33,f34,f35,f36,f37,f38,f39,f40,f41,f42,f43,f44,f45,f46,f47,f48,f49,f50,f51,f52,f53,f54,f55,f56,f57,f57a,f58b,f58,f59,f60,f61,f62,f63,f64,f65,f66,f67,f68}.
For comprehensive reviews see for example
\cite{review1,review2,review3,review4,f9,f12}. For recent work on
realistic GUT models see
\cite{vafa2,vafa3,review1,review2,review3,review4,uplift,f1,f2,f3,f4,f5,f6,f7,f8,f9,f10,f11,f12,f13,f14,f15,f16,f17,f18,f19,f21,f22,f23,f24,f25,f26,f27,f28,f29,f30,f31,f32,f33,f34,f35,f36,f37,f38,f39,f40,f41,f42,f43,f44,f45,f46,f47,f48,f49,f50,f51,f52,f53,f54,f55,f56,f57,f57a,f58b,f58,f59,f60,f61}.

In this work, we consider F-theory elliptic $K3$ fibrations over a
complex dimension two surface $S$. Particularly we focus our study
on the localized fields that are generated on the intersection
curve $\Sigma$ of two D7-branes in the absence of external gauge
fluxes. As we shall see, the fields that have localized solutions
along a matter curve $\Sigma$, are connected to an $N=2$
supersymmetric quantum mechanics algebra \cite{susyqm,susyqm1}. We
relate the number of the zero modes with the Witten index of the
susy algebra.

This paper is organized as follows: In section 1 we describe the
F-theory setup, that is, D7-branes intersections, matter curves
and the eight dimensional Super-Yang-Mills (SYM) theory. In
section 2 we give in brief, a self-contained review of the
supersymmetric quantum mechanics algebra. In section 3 we connect
the localized solutions of the BPS equations of motion, with an
$N=2$ supersymmetric quantum mechanics algebra, and relate the
total number of solutions to the Witten index. In section 4 we
generalize the latter to the case of three matter curves. Finally
in section 5 we present the conclusions.

\section{F-theory, D7-Branes Intersections and 8-dimensional SYM}

Singularities on complex manifolds play an important role in
string theory phenomenology
\cite{vafa1,vafa2,review1,review2,review3,review4,uplift,f1}.
Actually exceptional gauge groups are realized by the geometry of
singularities. For instance, real codimension four theories with
$N=1$ supersymmetry descend from M-theory defined on a seven
dimensional manifold with $G_2$ holonomy \cite{vafa2,review4,f1}.
The compactified wrapped space is $C^2/\Gamma$, with the last
being a singularity. Within the F-theory setup, $N=1$, $D=4$
supersymmetric gauge theories arise when F-theory is compactified
on Calabi-Yau fourfolds \cite{origins2,vafa2,review4,f1}. The
geometric techniques that can be defined on complex manifolds
enable us to build theories with rich phenomenology
\cite{vafa2,vafa3,review1,review2,review3,review4,uplift,f1,f2,f3,f4,f5,f6,f7,f8,f9,f10,f11,f12,f13,f14,f15,f16,f17,f18,f19,f21,f22,f23,f24,f25,f26,f27,f28,f29,f30,f31,f32,f33,f34,f35,f36,f37,f38,f39,f40,f41,f42,f43,f44,f45,f46,f47,f48,f49,f50,f51,f52,f53,f54,f55,f56,f57,f58,f59,f60,f61}.

\noindent We shall take into account F-theory compactifications on
Calabi-Yau fourfold that is actually an elliptic $K3$ fibration
over a complex dimension two surface $S$. Locally the theory is
described by the worldvolume of a D7 brane of ADE type which wraps
\linebreak $R^{1,3}\times S$ over the Calabi-Yau fourfold. The
K\"{a}hler surface $S$ is wrapped by the seven brane and it's
geometry can shrink to zero size inside a threefold base so we can
assume that gravity decouples from the theory. The resulting
theory in four dimensions is an $N=1$ supersymmetric theory
\cite{vafa2,f1}. Our analysis is based on references
\cite{vafa2,f1}.

\noindent The gauge group $G_S$ in F-theory is generated by
codimension one singularities which are the world volume of
D7-branes. The D7-branes wrap a complex dimension two surface $S$.
Let $H_S$ be the gauge group of a supersymmetric gauge field on
$S$, and also suppose \linebreak $H_S\subset G_S$. This gauge
field configuration breaks the gauge group $G_S$ down to $H_S$,
thus making possible to reduce the symmetries down, close to the
(usual) GUT ones. Specifically when the gauge group is of the form
$H_{S'}\times U(1)$, the resulting spectrum contains chiral matter
that come from the zero modes of bulk fields of the surface $S$.
Of course the way to generate the $H_S$ supersymmetric gauge
configuration is to give a vacuum expectation value to an
appropriate adjoint scalar of $G_S$.

\noindent Gauge groups are realized as ADE singularities on the
complex surface $S$. These singularities can be enhanced over
complex curves $\Sigma$ on the complex surfaces $S$. The curves
$\Sigma$ are actually Riemann surfaces and arise as the
intersection locus of the surface $S$ and another surface $S'$. As
a result bifundamental matter fields are defined on the
intersection $\Sigma$. Moreover a non-trivial background gauge
field on $\Sigma$ can induce chiral four dimensional matter.

\noindent We now describe in brief the field configurations that
reside in the complex surface $S$ \cite{vafa2,f1}. The physics of
the D7-branes wrapping $S$ is given in terms of an $D=8$ twisted
Super Yang-Mills on $R^{3,1}\times S$. The supersymmetric
multiplets include the gauge field plus a complex scalar $\varphi$
and the adjoint fermions $\eta ,\psi , \chi $. Let ($z_1,z_2$) be
the local coordinates that parameterize complex surface $S$. Then
the field content of the supermultiplets is:
\begin{equation}\label{gauge1}
A=A_{\mu}\mathrm{d}x^{\mu}+A_{m}\mathrm{d}z^{m}+A_{\bar{m}}\mathrm{d}\bar{z}^{m},{\,}{\,}{\,}{\,}\varphi
=\varphi_{12}{\,}\mathrm{d}z^1\wedge \mathrm{d}z^2
\end{equation}
and also,
\begin{equation}\label{gauge2}
\psi_{a}=\psi_{a\bar{1}}\mathrm{d}\bar{z}^1+\psi_{a\bar{2}}\mathrm{d}\bar{z}^2,{\,}{\,}{\,}{\,}\chi_a
=\chi_{a12}{\,}\mathrm{d}z^1\wedge \mathrm{d}z^2
\end{equation}
In the above, $\psi$ is a $(0,1)$-form while $\varphi$ and $\chi$
are $(2,0)$-forms (we dropped the subscript $a$ for simplicity).
The fermion $\eta$ is a $(0,0)$-form. Also $a=1,2$ and $m=1,2$.

\noindent The $N=1$, $D=4$ supersymmetric theory consists of the
gauge multiplet $(A_{\mu},\eta)$ together with the chiral
multiplets $(A_{\bar{m}},\psi_{\bar{m}})$ and
$(\varphi_{12},\chi_{12})$, plus their complex conjugates.

\noindent The $D=8$ effective action can be integrated over the
compact surface $S$ thus obtaining the  $D=4$ multiplet dynamics. In
reference to the Yukawa couplings, the most interesting term is the
superpotential,
\begin{equation}\label{superpotential}
W=M_*^4\int_S\mathrm{Tr}(F_S^{(0,2)}\wedge
{\bf\Phi})=W=M_*^4\int_S\mathrm{Tr}(\partial {\bf A}\wedge {\bf \Phi
})+M_*^4\int_S\mathrm{Tr}({\bf A}\wedge {\bf A}\wedge {\bf \Phi})
\end{equation}
with $M_*$ the mass scale of the supergravity low energy limit of
F-theory. The chiral superfields ${\bf A}$ and ${\bf \Phi}$ have
components,
\begin{equation}\label{gauge3}
{\bf A}_{\bar{m}}=A_{\bar{m}}+\sqrt{2}{\,}\theta
\psi_{\bar{m}}+\cdots
\end{equation}
and also,
\begin{equation}\label{gauge4}
{\bf \Phi}_{\bar{m}}=\phi_{12}+\sqrt{2}{\,}\theta \chi_{12}+\cdots
\end{equation}
where $\cdots$ involves auxiliary fields. Only the $(0,2)$ component
of the superstrength appears in (\ref{gauge3}) and (\ref{gauge4}).
In order to address the zero mode problem, we must find the
equations of motion that stem from the $D=8$ effective action.
Omitting any kinetic terms, the bilinear in the fermions, part of
the action is equal to,
\begin{equation}\label{bilinear}
I_F=\int_{R^{1,3}\times S}\mathrm{d}x^4\mathrm{Tr}\Big{(}\chi
\wedge\partial_A\psi+2{\,}i\sqrt{2}{\,}\omega \wedge\partial_A\eta
\wedge \psi+\frac{1}{2}\psi \wedge [\varphi,\psi]+\sqrt{2}{\,}\eta
[\bar{\varphi},\chi]+\mathrm{h.c.}\Big{)}
\end{equation}
where $\omega$ is the fundamental K\"{a}hler form of the complex
surface $S$. The variations of $\eta$, $\psi$ and $\chi$, yield
the equations of motion \cite{vafa2,f1}:
\begin{align}
& \omega\wedge \partial_A \psi+\frac{i}{2}[\bar{\phi},\chi]=0 \\
\notag & \bar{\partial}_A\chi-2i\sqrt{2}\omega\wedge \partial
\eta-[\varphi,\psi]=0 \\ \notag &
\bar{\partial}_A\psi-\sqrt{2}{\,}[\bar{\varphi},\eta]=0
\end{align}
In regard to the bosonic fields, the field strength $F_s$ must
have vanishing $(2,0)$ and $(0,2)$ components and hence it must
satisfy the BPS condition,
\begin{equation}\label{BPS}
\omega \wedge F_S+\frac{i}{2}[\varphi,\bar{\varphi}]=0
\end{equation}
Additionally, the complex scalar field must satisfy the
holomorphicity condition $\bar{\partial_A}\varphi =0$.

\noindent The charged massless multiplets in $D=4$ (localized zero
modes) are specified by the vacuum expectation value of the adjoint
scalar $\varphi$ and also from the background gauge field. When
$\langle \varphi \rangle =0$, the equations of motion imply that the
number of zero modes of $\psi$ and $\chi$ are determined by
topological invariants that depend both on $S$ and the background
gauge bundle.

\noindent In the next section we briefly present the supersymmetric
quantum mechanics algebra which we shall frequently use in the
subsequent sections.

\section{Supersymmetric Quantum Mechanics}

We review some issues related to supersymmetric quantum mechanics
\cite{susyqm,susyqm1} relevant to our analysis. Consider a quantum
system, described by a Hamiltonian $H$and characterized by the set
$\{H,Q_1,...,Q_N\}$, with $Q_i$ self-adjoint operators. The quantum
system is called supersymmetric, if,
\begin{equation}\label{susy1}
\{Q_i,Q_j\}=H\delta_{i{\,}j}
\end{equation}
with $i=1,2,...N$. The $Q_i$  are called supercharges and the
Hamiltonian ``$H$" is called SUSY Hamiltonian. The algebra
(\ref{susy1}) constitutes the so called N-extended supersymmetry.
Due to the anti-commutativity one has,
\begin{equation}\label{susy3}
H=2Q_1^2=Q_2^2=\ldots =2Q_N^2=\frac{2}{N}\sum_{i=1}^{N}Q_i^2.
\end{equation}
A supersymmetric quantum system $\{H,Q_1,...,Q_N\}$ is said to have
``good susy" (unbroken supersymmetry), if its ground state vanishes,
that is $E_0=0$. In the case $E_0>0$, that is for a positive ground
state energy, susy is said to be broken.

For good supersymmetry, the Hilbert space eigenstates must be
annihilated by the supercharges,
\begin{equation}\label{s1}
Q_i |\psi_0^j\rangle=0
\end{equation}
for all $i,j$.

\noindent In this paper we shall use the $N=2$ supersymmetric
quantum mechanics algebra. For convenience we shall refer to it as
``$N=2$ SUSY QM", or simply ``SUSY QM". We present in brief the
basic features of it. The $N=2$ algebra consists of two
supercharges $Q_1$ and $Q_2$ and a Hamiltonian $H$, which obey the
following relations,
\begin{equation}\label{sxer2}
\{Q_1,Q_2\}=0,{\,}{\,}{\,}H=2Q_1^2=2Q_2^2=Q_1^2+Q_2^2
\end{equation}
It is convenient for the our purposes to introduce the operator,
\begin{equation}\label{s2}
Q=\frac{1}{\sqrt{2}}(Q_{1}+iQ_{2})
\end{equation}
and the adjoint,
\begin{equation}\label{s255}
Q^{\dag}=\frac{1}{\sqrt{2}}(Q_{1}-iQ_{2})
\end{equation}
The above two satisfy the following equations,
\begin{equation}\label{s23}
Q^{2}={Q^{\dag}}^2=0
\end{equation}
and also are related to the Hamiltonian as,
\begin{equation}\label{s4}
\{Q,Q^{\dag}\}=H
\end{equation}
The Witten parity, $W$ for a $N=2$ algebra is defined as,
\begin{equation}\label{s45}
[W,H]=0
\end{equation}
and
\begin{equation}\label{s5}
\{W,Q\}=\{W,Q^{\dag}\}=0
\end{equation}
Also $W$ satisfies,
\begin{equation}\label{s6}
W^{2}=1
\end{equation}
By using $W$, we can span the Hilbert space $\mathcal{H}$ of the
quantum system to positive and negative Witten parity spaces. The
last are defined as,
$\mathcal{H}^{\pm}=P^{\pm}\mathcal{H}=\{|\psi\rangle :
W|\psi\rangle=\pm |\psi\rangle $. Therefore the quantum system
Hilbert space $\mathcal{H}$ is decomposed into the eigenspaces of
$W$, hence $\mathcal{H}=\mathcal{H}^+\oplus \mathcal{H}^-$.  Each
operator acting on the vectors of $\mathcal{H}$ can be represented
by $2N\times 2N$ matrices. We use the representation:
\begin{equation}\label{s7345}
W=\bigg{(}\begin{array}{ccc}
  I & 0 \\
  0 & -I  \\
\end{array}\bigg{)}
\end{equation}
with $I$ the $N\times N$ identity matrix. Recall that $Q^2=0$ and
$\{Q,W\}=0$, hence the supercharges are of the form,
\begin{equation}\label{s7}
Q=\bigg{(}\begin{array}{ccc}
  0 & A \\
  0 & 0  \\
\end{array}\bigg{)}
\end{equation}
and
\begin{equation}\label{s8}
Q^{\dag}=\bigg{(}\begin{array}{ccc}
  0 & 0 \\
  A^{\dag} & 0  \\
\end{array}\bigg{)}
\end{equation}
which imply,
\begin{equation}\label{s89}
Q_1=\frac{1}{\sqrt{2}}\bigg{(}\begin{array}{ccc}
  0 & A \\
  A^{\dag} & 0  \\
\end{array}\bigg{)}
\end{equation}
and also,
\begin{equation}\label{s10}
Q_2=\frac{i}{\sqrt{2}}\bigg{(}\begin{array}{ccc}
  0 & -A \\
  A^{\dag} & 0  \\
\end{array}\bigg{)}
\end{equation}
The $N\times N$ matrices $A$ and $A^{\dag}$, are generalized
annihilation and creation operators. The action of $A$ is defined
as $A: \mathcal{H}^-\rightarrow \mathcal{H}^+$ and that of
$A^{\dag}$ as, $A^{\dag}: \mathcal{H}^+\rightarrow \mathcal{H}^-$.
In the representation (\ref{s7345}), (\ref{s7}), (\ref{s8}) the
quantum mechanical Hamiltonian $H$, can be cast in a diagonal
form,
\begin{equation}\label{s11}
H=\bigg{(}\begin{array}{ccc}
  AA^{\dag} & 0 \\
  0 & A^{\dag}A  \\
\end{array}\bigg{)}
\end{equation}
Therefore for a $N=2$ supersymmetric quantum system, the total
supersymmetric Hamiltonian $H$, consists of two superpartner
Hamiltonians,
\begin{equation}\label{h1}
H_{+}=A{\,}A^{\dag},{\,}{\,}{\,}{\,}{\,}{\,}{\,}H_{-}=A^{\dag}{\,}A
\end{equation}
We can define an operator $P^{\pm}$. The eigenstates of $P^{\pm}$,
denoted as $|\psi^{\pm}\rangle$  are called positive and negative
parity eigenstates. These satisfy,
\begin{equation}\label{fd1}
P^{\pm}|\psi^{\pm}\rangle =\pm |\psi^{\pm}\rangle
\end{equation}
Using the representation (\ref{s7345}), the parity eigenstates are
represented in the form,
\begin{equation}\label{phi5}
|\psi^{+}\rangle =\left(%
\begin{array}{c}
  |\phi^{+}\rangle \\
  0 \\
\end{array}%
\right)
\end{equation}
and also,
\begin{equation}\label{phi6}
|\psi^{-}\rangle =\left(%
\begin{array}{c}
  0 \\
  |\phi^{-}\rangle \\
\end{array}%
\right)
\end{equation}
with $|\phi^{\pm}\rangle$ $\epsilon$ $H^{\pm}$.

\noindent ``Good supersymmetry" has some implications on the ground
states. For ``good supersymmetry", there must be at least one state
in the Hilbert space with vanishing energy eigenvalue, that is
$H|\psi_{0}\rangle =0$. The Hamiltonian commutes with the
supercharges, $Q$ and $Q^{\dag}$, hence it is obvious that,
$Q|\psi_{0}\rangle =0$ and $Q^{\dag}|\psi_{0}\rangle =0$. For a
ground state with negative parity,
\begin{equation}\label{phi5}
|\psi^{-}_0\rangle =\left(%
\begin{array}{c}
  |\phi^{-}_{0}\rangle \\
  0 \\
\end{array}%
\right)
\end{equation}
this would imply that $A|\phi^{-}_{0}\rangle =0$, while for a
positive parity ground state,
\begin{equation}\label{phi6s6}
|\psi^{+}_{0}\rangle =\left(%
\begin{array}{c}
  0 \\
  |\phi^{+}_0\rangle \\
\end{array}%
\right)
\end{equation}
it would imply that $A^{\dag}|\phi^{+}_{0}\rangle =0$. Generally a
ground state can have either positive or negative Witten parity.
In the case the ground state is degenerate both cases can occur.
When it happens $E\neq 0$, the number of positive parity
eigenstates is equal to the negative parity eigenstates. This does
not occur for the zero modes. A rule to decide if there are zero
modes is given by the Witten index. Let $n_{\pm}$ the number of
zero modes of $H_{\pm}$ in the subspace $\mathcal{H}^{\pm}$. For
finite $n_{+}$ and $n_{-}$ the quantity,
\begin{equation}\label{phil}
\Delta =n_{-}-n_{+}
\end{equation}
is called the Witten index. When the Witten index is non-zero
integer, supersymmetry is good-unbroken. If the Witten index is
zero, it is not clear whether supersymmetry is broken (which would
mean $n_{+}=n_{-}=0$) or not ($n_{+}= n_{-}\neq 0$).

\noindent The Fredholm index of the operator $A$ defined as,
\begin{equation}\label{ker}
\mathrm{ind} A = \mathrm{dim}{\,}\mathrm{ker}
A-\mathrm{dim}{\,}\mathrm{ker} A^{\dag}=
\mathrm{dim}{\,}\mathrm{ker}A^{\dag}A-\mathrm{dim}{\,}\mathrm{ker}AA^{\dag}
\end{equation}
is obviously related to the Witten index as follows,
\begin{equation}\label{ker1}
\Delta=\mathrm{ind} A=\mathrm{dim}{\,}\mathrm{ker}
H_{-}-\mathrm{dim}{\,}\mathrm{ker} H_{+}
\end{equation}
The Fredholm index is a topological invariant and we shall only
consider Fredholm operators.

\section{Fields at D7-branes Intersections}

As we mentioned in the previous sections, the localized fields are
on a matter curve $\Sigma$ which is the intersection of the
complex surfaces $S$ and $S'$. The preservation of $N=1$
supersymmetry in $D=4$, requires the theory (that actually
describes matter) defined on $R^{1,3}\times \Sigma$ to be $D=6$
twisted super Yang-Mills \cite{vafa2,f1}. The six dimensional
twisted supermultiplets include two complex scalars and a Weyl
spinor. In four dimensions this decomposes into the dimension four
chiral multiplets $(\sigma,\lambda)$ and $(\sigma^c,\lambda^c)$
plus the $CPT$-conjugate. The number of zero modes is given by
topological invariants inherent to the curve $\Sigma$ and the
gauge bundle in $G_{\Sigma}$ (we shall say more on this, at the
conclusion section). In this section we shall relate the number of
zero modes to the Witten index of an $N=2$ SUSY QM algebra. Let
the $D=8$ theory on $S$ have a gauge group $G_{\Sigma}$. We give a
coordinate dependent vacuum expectation value to the adjoint
scalar, of the form,
\begin{equation}\label{backscalar}
\langle \varphi \rangle = m^2z_1Q_1
\end{equation}
with $z_1$ a complex coordinate on $S$, $Q_1$ a $U(1)$ generator
of the Cartan subalgebra of $G_{\Sigma}$ and
$\varphi=\varphi_{12}$. The parameter ``$m$" is a mass parameter
we used, in order $\varphi$ has the correct dimensions. When the
vacuum expectation value of $\varphi$ is non zero, the $D=8$
fields have zero modes localized at $z_1=0$ that are connected to
the fields appearing in $\Sigma$ \cite{vafa2,f1}. When $z_1=0$,
the gauge group is unbroken, but when $z_1\neq 0$, the group
breaks to,
\begin{equation}\label{subgropoup}
G_S\times U(1)\subset G_{\Sigma}
\end{equation}
and the generators of $G_S$ commute with $Q_1$. The locus $z_1=0$
defines the intersection curve $\Sigma$ and hence at $z_1=0$ the
symmetry enhances to $G_{\Sigma}$.

\noindent In reference to D7-branes, an adjoint scalar corresponds
to degrees of freedom in the transverse direction, and thus a non
zero vacuum expectation value means that some branes are separated
and the gauge group is broken. For example if a brane is moved from
a stack of $K+1$ D7-branes the original $SU(K+1)$ symmetry is broken
to $SU(K)\times U(1)$. This way, the resulting theory at the
intersection contains localized massless bifundamentals of the form
$(K,-1)\oplus (\bar{K},+1)$. The massless bifundamentals coming from
the adjoint of $G_{\Sigma}$ are classified in terms of the
irreducible representations of $G_S\times U(1)$. For interesting
examples of this see \cite{vafa2}.

\noindent In order to find how localized fermion matter on
$\Sigma$ results from zero modes of the $D=8$ bulk theory, we must
solve the $D=8$ equations of motion for the twisted fermions. The
assumptions are that $\varphi$, has a non-trivial $z_1$-dependent
vacuum expectation value, like in (\ref{backscalar}) and that no
background gauge fields (usually $U(1)$'s) are present. The
localized fermions correspond to the $\lambda$ and $\lambda^{c}$
fermions of the twisted Super Yang-Mills on $R^{1,3}\times
\Sigma$. Let the K\"{a}hler form of $S$ be of the form,
\begin{equation}\label{kaeheregut}
\omega =\frac{i}{2}\large{(}\mathrm{d}z^1\wedge
\mathrm{d}\bar{z}^1+\mathrm{d}z^2\wedge \mathrm{d}\bar{z}^2\large{)}
\end{equation}
The coordinates of $S$, $z_1$ and $z_2$, parameterize the
intersection $\Sigma$ in transverse and tangent directions
respectively. With $\omega$ being of the form (\ref{kaeheregut}) and
neglecting the $z_2$ derivatives, the equations of motion can be
cast as \cite{vafa2,f1}:
\begin{align}\label{eqmotion1}
& \sqrt{2}\partial_1\eta
-m^2z_1q_1\psi_{\bar{2}}=0{\,}{\,}{\,}{\,}{\,}{\,}{\,}{\,}{\,}{\,}{\,}{\,}{\,}{\,}{\,}{\,}{\,}{\,}\partial_1
\psi_{\bar{1}}-m^2\bar{z}_1q_1\chi =0
\\  \notag & \partial_1\psi_{\bar{2}}-\sqrt{2}{\,}m^2z_1q_1\eta
=0
{\,}{\,}{\,}{\,}{\,}{\,}{\,}{\,}{\,}{\,}{\,}{\,}{\,}{\,}{\,}{\,}{\,}{\,}\bar{\partial}
        _1\chi-m^2z_1q_1\psi_{\bar{1}}=0
\end{align}
with $\chi =\chi_{12}$. In the above, $q_1$ stands for the $U(1)$
fermion charge belonging to a representation $(R,q_1)$ of
$G_S\times U(1)$. It is obvious that there are no localized
solutions for the fermions $\eta$ and $\psi_{\bar{2}}$ \cite{f1}.
On the contrary the solutions for $\chi$ and $\psi_{\bar{1}}$ are:
\begin{equation}\label{localizedsol}
\chi=f(z_2)e^{-q_1m^2|z_1|^2},{\,}{\,}{\,}\psi_{\bar{1}}=-f(z_2)e^{-q_1m^2|z_2|^2}.
\end{equation}
The solutions are peaked around $z_1=0$. Also the constant
$q_1m^2$ is of the order of the F-theory mass scale $M_*^2$. There
are similar equations of motion stemming from the hermitian
conjugate terms in relation (\ref{bilinear}) which we omitted for
simplicity. These act on the conjugate fermions $\bar{\psi}$ and
$\bar{\chi}$.

\noindent Let us see how the number of the zero modes
corresponding to the fields $\psi_{\bar{1}}$, $\chi$, is related
to the Witten index. We define,
\begin{equation}\label{dmatrix}
D=\left(%
\begin{array}{cc}
  \partial_1 & -m^2\bar{z}_1q_1 \\
  -m^2z_1q_1 & \bar{\partial}_{1} \\
\end{array}%
\right)
\end{equation}
and additionally,
\begin{equation}\label{dmatrix}
D^{\dag}=\left(%
\begin{array}{cc}
  \bar{\partial}_{1} & -m^2\bar{z}_1q_1 \\
  -m^2z_1q_1 & \partial_{1} \\
\end{array}%
\right)
\end{equation}
acting on,
\begin{equation}\label{wee}
\left(%
\begin{array}{c}
  \psi_{\bar{1}} \\
  \chi \\
\end{array}%
\right)
\end{equation}
The solutions of the equations of motion (\ref{eqmotion1}) for
$\psi$ and $\chi$ are the zero modes of $D$. The Fredholm index
$I_D$ of the operator $D$, is equal to,
\begin{equation}\label{indexd}
\mathrm{ind}_D=\mathrm{I}=\mathrm{dim{\,}ker}(D^{\dag})-\mathrm{dim{\,}ker}(D)
\end{equation}
which is the number of zero modes of $\mathcal{D}$ minus the
number of zero modes of $\mathcal{D}^{\dag}$.

\noindent Using $D$ we can define an $N=2$ supersymmetric quantum
mechanical system. Indeed we can write,
\begin{equation}\label{wit2}
Q=\bigg{(}\begin{array}{ccc}
  0 & D \\
  0 & 0  \\
\end{array}\bigg{)}
\end{equation}
and additionally,
\begin{equation}\label{wit3}
Q^{\dag}=\bigg{(}\begin{array}{ccc}
  0 & 0 \\
  D^{\dag} & 0  \\
\end{array}\bigg{)}
\end{equation}
Also the Hamiltonian of the system can be written,
\begin{equation}\label{wit4}
H=\bigg{(}\begin{array}{ccc}
  DD^{\dag} & 0 \\
  0 & D^{\dag}D  \\
\end{array}\bigg{)}
\end{equation}
It is obvious that the above matrices obey, $\{Q,Q^{\dag}\}=H$,
$Q^2=0$, ${Q^{\dag}}^2=0$, $\{Q,W\}=0$, $W^2=I$ and $[W,H]=0$.
Thus we can relate the Witten index of the $N=2$ supersymmetric
quantum mechanics system, to the index $I_D$ of the operator $D$.
Indeed we have $I_D=-\Delta$, because,
\begin{equation}\label{ker}
I_D=\mathrm{dim}{\,}\mathrm{ker}
D^{\dag}-\mathrm{dim}{\,}\mathrm{ker} D=
\mathrm{dim}{\,}\mathrm{ker}DD^{\dag}-\mathrm{dim}{\,}\mathrm{ker}D^{\dag}D=-\mathrm{ind}D=-\Delta=n_--n_+
\end{equation}
with $n_{-}$ and $n_+$ defined above equation (\ref{phil}). Owing
to the supersymmetric quantum mechanical structure of the system,
the zero modes of the operators $D$ and $D^{\dag}$ are related to
the zero modes of the operators $DD^{\dag}$ and $D^{\dag}D$. The
above imply that the zero modes of the operator $DD^{\dag}$ and
also of $D^{\dag}D$ can be classified according to the Witten
parity, to parity positive and parity negative solutions.

\noindent If the theory defined on $\Sigma$ can be viewed as a
cosmic string defect on $S$ \cite{vafa2}, the classification of
states in parity even and parity odd states, could be useful. This
is due to the fact that, if someone could define a movement of the
localized modes, the classification in parity even and odd states
could be an analogue of L-movers and R-movers of the
superconducting string case (of course this assumption could be
true if some restrictions are imposed on the function $f(z_2)$).
Actually GUT inspired fermionic zero modes in superconducting
strings backgrounds, can be classified to parity even and parity
odd states, owing to an underlying SUSY QM algebra
\cite{oikonomou}.

\section{Intersecting Matter Curves, Fermionic Zero Modes and SUSY QM}

Consider now three matter curves. These matter curves can
intersect at a point. We denote the three matter curves as
$\Sigma_i$, with $i=1,2,3$. Each matter curve has a group $G_i$,
that on the intersection point further enhances to a higher group
$G_p$. In order to find the localized fermionic solutions of the
eight dimensional theory on $S$ one must have a non-trivial
background for the adjoint scalar, as in the previous section.
Next, the solutions must be found for each matter curve
\cite{vafa2,f1}.

\noindent In the case we are studying, a more involved vacuum
expectation value is needed, which looks like \cite{vafa2,f1}:
\begin{equation}\label{backscalar1}
\langle \varphi \rangle = m^2z_1Q_1+m^2z_2Q_2
\end{equation}
In the above, $Q_1$ and $Q_2$ are the $U(1)$ generators that are
included in the enhancement group $G_p$ at the intersection point,
and ``$m_1$" and ``$m_2$" are mass scales that are related to the
F-theory mass scale $M_*$. Taking $m_1=m_2=m$ will simplify things
but will not change the results.

\noindent The above form of the vacuum expectation value of the
adjoint scalar field resolves the $G_p$ singularity at the
intersection point. We suppose the intersection point is
$(z_1,z_2)=(0,0)$. The three different curves $\Sigma_1$,
$\Sigma_2$, $\Sigma_3$ are defined by the loci $z_1=0$, $z_2=0$
and $z_1+z_2=0$ respectively. Each curve represents a fermion and
we can say that under the $U(1)$ charges, the curves are
classified according to the following table,
\begin{center}\label{table2}
\begin{tabular}{|c|c|c|}
  \hline
  \bf{matter curve} & ($q_1,q_2$) & \bf{locus} \\
  \hline
  $\Sigma_1$ & $(q_1,0)$  & $z_1=0$\\
  \hline
  $\Sigma_2$ & $(0,q_2)$ & $z_2=0$\\
  \hline
  $\Sigma_3$ & $(-q_1,-q_2)$ & $z_1+z_2${\,}={\,}0\\
  \hline
\end{tabular}
\\ \bigskip{ \bfseries{Classification of the matter curves}}
\end{center}
\bigskip

\noindent where the constants $(q_1,q_2)$ are the $U(1)$ charges
of the fermions belonging to an irreducible representation
$(R,q_1,q_2)$ of $G_S\times U(1)_1\times U(1)_2$ (note that $Q_1$
generates $U(1)_1$ and $Q_2$ generates $U(1)_2$). With the adjoint
vacuum expectation value being that of relation
(\ref{backscalar1}) the equations of motion are:
\begin{align}\label{eqmotion2}
& \partial_2\psi_{\bar{2}}+\partial_1\psi_{\bar{1}}
-m^2(\bar{z}_1q_1+\bar{z}_2q_2)\chi =0
\\  \notag & {\,}{\,}{\,}{\,}{\,}{\,}{\,}{\,}{\,}{\,}{\,}{\,}{\,}{\,}{\,}{\,}{\,}{\,}{\,}\bar{\partial}_1\chi-m^2(z_1q_1+z_2q_2)\psi_{\bar{1}} =0
\\  \notag & {\,}{\,}{\,}{\,}{\,}{\,}{\,}{\,}{\,}{\,}{\,}{\,}{\,}{\,}{\,}{\,}{\,}{\,}{\,}\bar{\partial}_2\chi-m^2(z_1q_1+z_2q_2)\psi_{\bar{2}} =0
\end{align}

\subsection{Localized fermion around $z_1=0$}

In the case of the curve $\Sigma_1$, we have $q_2=0$. The localized
fermions are at $z_1=0$. Localized solutions to the equations of
motion (\ref{eqmotion2}) are \cite{f1}:
\begin{equation}\label{locali1}
\psi_{\bar{2}}=0,{\,}{\,}{\,}{\,}{\,}{\,}\chi=f(z_2)e^{-q_1m^2|z_1|^2},{\,}{\,}{\,}\psi_{\bar{1}}=-\chi.
\end{equation}
with $f(z_2)$ a holomorphic function of $z_2$. We can associate a
$N=2$ SUSY QM algebra to this matter curve. Using the notation of
the previous section, we can define the matrix $D_1$ and also
$D^{\dag}_1$ as follows,
\begin{equation}\label{dmatrix1}
D_1=\left(%
\begin{array}{cc}
  \partial_1 & -m^2(\bar{z}_1q_1+\bar{z}_2q_2) \\
  -m^2(z_1q_1+z_2q_2) & \bar{\partial}_{1} \\
\end{array}%
\right)
\end{equation}
and,
\begin{equation}\label{dmatrix23}
D^{\dag}_1=\left(%
\begin{array}{cc}
  \bar{\partial}_{1} & -m^2(\bar{z}_1q_1+\bar{z}_2q_2) \\
  -m^2(z_1q_1+z_2q_2) & \partial_{1} \\
\end{array}%
\right)
\end{equation}
acting on,
\begin{equation}\label{wee33}
\left(%
\begin{array}{c}
  \psi_{\bar{1}} \\
  \chi \\
\end{array}%
\right)
\end{equation}
Then all the results of the previous section, apply in this case. We
must note that the SUSY QM structure exists if $\psi_{\bar{2}}=0$ on
this matter curve. In the converse case, we cannot define a matrix
like $D_1$.

\subsection{Localized fermion around $z_2=0$}

In the case of the curve $\Sigma_2$, we have $q_1=0$. The localized
fermions are at $z_2=0$. Localized solutions to the equations of
motion (\ref{eqmotion2}) are:
\begin{equation}\label{locali2}
\psi_{\bar{2}}=-\chi,{\,}{\,}{\,}{\,}{\,}{\,}\chi=g(z_2)e^{-q_2m^2|z_1|^2},{\,}{\,}{\,}\psi_{\bar{1}}=0.
\end{equation}
with $g(z_1)$ a holomorphic function of $z_1$. Thus the $N=2$ SUSY
QM algebra can be defined in terms of the $D_2$ matrix, which is
equal to:
\begin{equation}\label{dmatrix2344}
D_2=\left(%
\begin{array}{cc}
  \partial_2 & -m^2(\bar{z}_1q_1+\bar{z}_2q_2) \\
  -m^2(z_1q_1+z_2q_2) & \bar{\partial}_{2} \\
\end{array}%
\right)
\end{equation}

\subsection{Localized fermion around $z_1+z_2=0$}

The matter curve $\Sigma_3$, has generic charges $q_1$ and $q_2$.
To make things easier we make the following transformations:
\begin{align}\label{transforma}
&w=z_1+z_2,{\,}{\,}{\,}{\,}{\,}{\,}{\,}{\,}{\,}{\,}{\,}{\,}\psi_{\bar{w}}=\frac{1}{2}\large{(}\psi_{\bar{1}}+\psi_{\bar{2}}\large{)}
\\ & \notag u=z_1-z_2,{\,}{\,}{\,}{\,}{\,}{\,}{\,}{\,}{\,}{\,}{\,}{\,}\psi_{\bar{u}}=\frac{1}{2}\large{(}\psi_{\bar{1}}-\psi_{\bar{2}}\large{)}
\end{align}
Then the equations of motion (\ref{eqmotion2}) can be cast as:
\begin{align}\label{eqmotion3}
& 2{\,}\partial_w\psi_{\bar{w}}+2{\,}\partial_u\psi_{\bar{u}}
-\frac{m^2}{2}\big{(}\bar{w}(q_1+q_2)+\bar{u}(q_1-q_2)\big{)}\chi =0
\\  \notag & {\,}{\,}{\,}{\,}{\,}{\,}{\,}{\,}{\,}{\,}{\,}{\,}{\,}{\,}{\,}{\,}{\,}{\,}{\,}{\,}{\,}{\,}{\,}{\,}{\,}2{\,}\bar{\partial}_{\bar{w}}\chi-m^2\Big{(}w(q_1+q_2)+u(q_1-q_2)\Big{)}\psi_{\bar{w}} =0
\\  \notag & {\,}{\,}{\,}{\,}{\,}{\,}{\,}{\,}{\,}{\,}{\,}{\,}{\,}{\,}{\,}{\,}{\,}{\,}{\,}{\,}{\,}{\,}{\,}{\,}{\,}{\,}2{\,}\bar{\partial}_{\bar{u}}\chi-m^2\Big{(}w(q_1+q_2)+u(q_1-q_2)\Big{)}\psi_{\bar{u}} =0
\end{align}
In the same way as in the two previous cases a $N=2$ SUSY QM algebra
underlies the fermion system when $\psi_{\bar{u}}=0$. Indeed we can
define the matrices $D_3$ and $D^{\dag}_3$ as:
\begin{equation}\label{dmat}
D_3=\left(%
\begin{array}{cc}
  2{\,}\partial_w & -\frac{m^2}{2}\Big{(}\bar{w}(q_1+q_2)+\bar{u}(q_1-q_2)\Big{)} \\
  -m^2\Big{(}w(q_1+q_2)+u(q_1-q_2)\Big{)} & 2{\,}\bar{\partial}_w \\
\end{array}%
\right)
\end{equation}
and,
\begin{equation}\label{dmatr}
D^{\dag}_3=\left(%
\begin{array}{cc}
  2{\,}\bar{\partial}_w  & -m^2\Big{(}\bar{w}(q_1+q_2)+\bar{u}(q_1-q_2)\Big{)} \\
  -\frac{m^2}{2}\Big{(}w(q_1+q_2)+u(q_1-q_2)\Big{)} & 2{\,}\partial_w \\
\end{array}%
\right)
\end{equation}
acting on,
\begin{equation}\label{we23e}
\left(%
\begin{array}{c}
  \psi_{\bar{w}} \\
  \chi \\
\end{array}%
\right)
\end{equation}
The localized solutions to the equations of motion
(\ref{eqmotion3}) around $z_1+z_2=0$ are:
\begin{equation}\label{locali3}
\psi_{\bar{w}}=\frac{1}{\sqrt{2}}\chi,{\,}{\,}{\,}{\,}{\,}{\,}\chi=g(u)e^{-\frac{q_2m^2}{\sqrt{2}}|w|^2},{\,}{\,}{\,}\psi_{\bar{u}}=0.
\end{equation}
In conclusion each matter curve corresponds to an underlying $N=2$
SUSY QM algebra that can be built using the three matrices $D_1$,
$D_2$ and $D_3$. In addition the zero modes of $D_1$, $D_2$ and
$D_3$ correspond to the solutions of (\ref{eqmotion2}).

Before closing we must note that an $N=2$ supersymmetric structure
was expected in the system since, the dimensional reduction of the
intersection of two D7-branes gives rise to a four-dimensional
$N=2$ sector of the full effective field theory.

\section{Conclusions}

In this article we have connected the system of fermion matter
fields that exist at the intersection locus of D7-branes, with a
supersymmetric quantum mechanics algebra, within the theoretical
framework of F-theory. We found that only the localized fields
along the complex dimension one intersection (Riemann surface) are
related to the SUSY QM algebra, and we examined three
(phenomenologically important) matter curves and their localized
solutions. The zero modes solutions of the equations of motion are
related to the Witten index of the underlying SUSY QM algebra.

\noindent Mention that in order to obtain non-trivial fermion mass
hierarchies, non-constant fluxes are needed. We have not checked
whether there is an underlying $N=2$ SUSY QM algebra in the
presence of non-constant and also constant fluxes.  An interesting
case, although we shall not pursuit here.

Before closing, we must mention that there are cohomological
techniques that determine the massless spectrum localized on the
intersection $\Sigma$ \cite{vafa2}. Indeed (following
\cite{vafa2}) if $\mathcal{H}$ denotes the group that remains at
the intersection $\Sigma$, $\mathcal{V}_i$ the gauge bundle that
transforms as a representation of $\mathcal{H}$ and $K_{\Sigma}$
the canonical bundle on $\Sigma$, the net chirality of the zero
modes is \cite{vafa2}:
\begin{equation}\label{netchirallity}
n_{\nu_{i}}-n_{\nu_{i}}^*=(1-g)\mathrm{rk}\large{(}K_{\Sigma}^{1/2}\otimes
\mathcal{V}_i\large{)}+\int_{\Sigma}c_1\large{(}K_{\Sigma}^{1/2}\otimes
\mathcal{V}_i\large{)}
\end{equation}
In the above equation, $n_{\nu_{i}}$ and $n_{\nu_{i}}^*$ stand for
the number of generations and anti-generations respectively, while
$g$ is the genus of the curve $\Sigma$.

\section*{Acknowledgments}

V. Oikonomou is indebted to Prof. G. Leontaris for useful
discussions on F-theory GUTs and related issues.

\end{document}